# Discrete-time Calogero-Moser Model and Lattice KP Equations

F. W. Nijhoff and G. D. Pang

ABSTRACT. We introduce an integrable time-discretized version of the classical Calogero-Moser model, which goes to the original model in a continuum limit. This discrete model is obtained from pole solutions of a semi-discretized version of the Kadomtsev-Petviashvili equation, leading to a finite-dimensional symplectic mapping. Lax pair, symplectic structure and sufficient set of invariants of the discrete Calogero-Moser model are constructed for both the rational and elliptic cases. The classical $r$-matrix is the same as for the continuum model. An exact solution of the initial value problem is given for the rational discrete-time Calogero-Moser model. The pole-expansion and elliptic solutions of the fully discretized Kadomtsev-Petviashvili equation are also discussed.

## 1. Introduction

The classical Calogero-Moser (CM) model, [1, 2], is a one-dimensional many-body system with long-range interactions, and it is described by the equations of motion

$$(1.1) \qquad \ddot{x}_i = g \sum_{\substack{j=1 \\ j \neq i}}^{N} \wp'(x_i - x_j) , \qquad i = 1, 2, ..., N,$$

in which $\wp(x) = \wp(x|\omega_1, \omega_2)$ is the Weierstrass elliptic P-function with periods $\omega_1$ and $\omega_2$. In the degenerate cases when one or both of the periods $\omega_{1,2}$ take infinite values, $\wp(x)$ becomes: I) $x^{-2}$ when $\omega_1 \to \infty$ and $\omega_2 \to i\infty$ (This is called the rational case, and the eq. (1.1) in which $\wp(x)$ is replaced by $x^{-2}$ will be called the rational CM model in the following); II) $(sinx)^{-2} - \frac{1}{3}$ when $\omega_1 = \pi/2$ and $\omega_2 \to i\infty$ (the trigonometric case); III) $(sinhx)^{-2} + \frac{1}{3}$ when $\omega_1 \to \infty$ and $\omega_2 = i\pi/2$ ( the hyperbolic case). This classical CM model, as well as its quantum analogue, are integrable systems, cf. the two review papers, [3], for a comprehensive account of both cases. Integrable long-range interaction models of CM type, and their relativistic deformations, [4], have gained new interest recently in connection with the representation theory of Lie algebras and affine quantum algebras, cf. e.g. [5, 6]. Furthermore, there is strong evidence that the CM model plays an important role in understanding the universal behaviour in quantum chaos, [7]. As other longe-range integrable models with similar connections, we can mention the Haldane-Shastry SU(N) chain, cf. [8], the Gaudin model, [9], and the Garnier system, [10].

1991 *Mathematics Subject Classification*. Primary 58F07, 39A10; Secondary 70F10.
Both of the authors were supported by the Alexander von Humboldt Foundation.

In a recent article, [**11**], we introduced an exact (time-)discretization of the classical CM model in the rational case. We know that many of the conventional continuous integrable models have exactly integrable discrete-time counterparts (see e.g. some of the other contributions at this Workshop), although it is often highly nontrivial to obtain them. These exact time-discretizations have a great potential for applications (numerical integration schemes, construction of cellular automata, discrete "differential" geometry, physical modelling). They provide exact interpolations and thereby finite algorithms for the calculation of classical orbits or –in the quantum regime– of quantum-mechanical propagators. As such, having exact discrete-time analogue of the CM model could be of use in the random matrix approach to quantum chaos. From a more general point of view, such a discrete-time long-range model might exhibit some new features, from which much can still be learnt about the true nature of integrability in the discrete regime.

In this paper we will review the construction of the rational (integrable) discrete-time CM model and also generalize this discrete model to the elliptic case, namely we will construct an elliptic (integrable) discrete-time CM model, which goes exactly to (1.1) in a continuum limit. The pole-expansion and elliptic solutions of the lattice Kadomtsev-Petviashvili (KP) equation will also be discussed in this paper.

## 2. The rational discrete-time CM model

The method used here to obtain the discrete-time CM model is based on the observation of [**12**], cf. also [**13**], that the dynamics of the poles of special solutions of integrable nonlinear evolution equations is connected with integrable systems of particles on the line. The connection between the pole solutions of the (continuum) KP equation and the CM system was found by Krichever in [**14**], cf. also [**15**]. Here we will perform a similar construction for the discrete case. We will first show that the rational pole solutions of a semi-discretized version of the KP equation is connected with a time-discretized version of the rational CM model. Later in this paper we will generalize this result to the elliptic case and we will also discuss on a fully-discretized version of the KP equation, i.e., the lattice KP equation.

The semi-discretized version of the KP equation reads

$$(2.1) \qquad (p - q + \hat{u} - \widetilde{u})_\xi = (p - q + \hat{u} - \widetilde{u})(u + \widetilde{\hat{u}} - \hat{u} - \widetilde{u}),$$

where $p$ and $q$ are two (lattice) parameters, $u$ is the (classical) field, the $\sim$ denotes the discrete time-shift corresponding to a translation in the "time" direction while the $\char94$ denotes a shift or translation in the "spatial" direction. In (2.1) the time and spatial variables are discrete while the third variable $\xi$ is continuous. Eq. (2.1) is obtained from the fully-discretized version of the KP equation of [**16**] (see also (5.1) below) by letting one of three lattice parameters tend to zero.

It can be easily proved that the compatibility condition of the two equations

$$(2.2) \qquad \widetilde{\phi} = \phi_\xi + (p + u - \widetilde{u})\phi,$$
$$(2.3) \qquad \hat{\phi} = \phi_\xi + (q + u - \hat{u})\phi,$$

leads to (2.1), and (2.2-2.3) is called the Lax representation of (2.1).

Direct calculation shows that

$$(2.4) \qquad u = \frac{1}{x} \quad \text{with} \quad x = \frac{n}{p} + \frac{m}{q} + \xi$$

is a simple rational (pole) solution to the semi-discretized version of KP equation (2.1). Here $n(m)$ is an integer, which is increased by value 1 when operated by $\widetilde{\phantom{x}}$ ($\widehat{\phantom{x}}$). If we substitute this solution (2.4) into the $u$'s in (2.2-2.3), then we find that

$$(2.5) \qquad \phi = \left(1 - \frac{1}{kx}\right)(p+k)^n(q+k)^m exp(k\xi)$$

satisfies (2.2-2.3), where $k$ is a spectral parameter.

Enlightened by the above simple exercise, we now suppose that

$$(2.6) \qquad u = \sum_{i=1}^{N} \frac{1}{\xi - x_i},$$

$$(2.7) \qquad \phi = \left(1 - \frac{1}{k}\sum_{i=1}^{N}\frac{b_i}{\xi - x_i}\right)(p+k)^n(q+k)^m exp(k\xi),$$

where $x_i$ and $b_i$ are independent of $\xi$ (but they depend on the time variable), and we are to find the conditions these $x_i$ and $b_i$ should satisfy such that equation (2.2) is valid. Substituting (2.6-2.7) into (2.2) and equating to zero the coefficients of $(\xi - x_i)^{-1}$ and $(\xi - \widetilde{x}_i)^{-1}$, we obtain the following equations:

$$(2.8) \qquad (p+k)b = ke + Lb,$$
$$(2.9) \qquad (p+k)\widetilde{b} = ke + Mb,$$

where the vectors $b = (b_1, b_2, ..., b_N)^T$ and $e = (1, 1, 1, ..., 1)^T$ and the matrices

$$(2.10) \qquad L = \sum_{i,j=1}^{N} \frac{e_{ii}}{x_i - \widetilde{x}_j} - \sum_{\substack{i,j=1 \\ i\neq j}}^{N} \frac{e_{ii} + e_{ij}}{x_i - x_j},$$

$$(2.11) \qquad M = \sum_{i,j=1}^{N} \frac{e_{ji}}{x_i - \widetilde{x}_j}.$$

In the above equations, $\widetilde{b}$ and $\widetilde{x}_i$ denotes the discrete time-shift as stated before, and the elements of the $N \times N$ matrices $e_{ij}$ are defined by $(e_{ij})_{kl} = \delta_{ik}\delta_{jl}$. The compatibility of (2.8-2.9) leads to the equation

$$(2.12) \qquad (\widetilde{L}M - ML)b + k(\widetilde{L} - M)e = 0.$$

(2.12) is a discrete *inhomogeneous* Lax's equation. It can be readily checked that the resulting equations of (2.12), i.e.

$$(2.13) \qquad \widetilde{L}M = ML$$

and

$$(2.14) \qquad (\widetilde{L} - M)e = 0,$$

are consistent and give the *same* discrete equations of motion of a $N-$particle system:

$$(2.15) \quad \frac{1}{x_i - \widetilde{x}_i} + \frac{1}{x_i - \underset{\sim}{x}_i} + \sum_{\substack{j=1 \\ j \neq i}}^{N} \left( \frac{1}{x_i - \widetilde{x}_j} + \frac{1}{x_i - \underset{\sim}{x}_j} - 2\frac{1}{x_i - x_j} \right) = 0,$$

$$i = 1, 2, ..., N,$$

where $\underset{\sim}{x}_i$ denotes the discrete time-shift in the reverse direction to the one of $\widetilde{x}_i$. We will call the model, for which eq. (2.15) are the equations of motion, the rational discrete-time CM model. We will show below that in a continuum limit these equations go to that of the original (rational) CM model.

The Lax pair for the rational discrete-time CM model is given by $L$ and $M$ in (2.10-2.11). As a consequence of eq. (2.13) we have

$$(2.16) \qquad \widetilde{I}_k = I_k$$

for any $k = 1, 2, ...$, where $I_k \equiv Tr(L^k)$. Thus the construction of invariants of the discrete-time flow given by (2.15) is similar to the continuous case.

## 3. Integrability, integration and continuum limit

**3.1. Complete integrability.** In order to establish the complete integrability of the rational discrete CM model (2.15), we need first to establish an appropriate symplectic structure for the N-particle system. We start by noting that eq. (2.15) can actually be obtained from the variation of a discrete action, given by

$$(3.1) \quad S = \sum_n \mathcal{L}(x, \widetilde{x}) = \sum_n \left( -\sum_{i,j=1}^{N} \log|x_i - \widetilde{x}_j| + \sum_{\substack{i,j=1 \\ i \neq j}}^{N} \log|x_i - x_j| \right),$$

in which the sum over $n$ denotes the sum over all discrete-time iterates. The discrete Euler-Lagrange equations

$$(3.2) \qquad \widetilde{\frac{\partial \mathcal{L}}{\partial x_i}} + \frac{\partial \mathcal{L}}{\partial \widetilde{x}_i} = 0, \quad i = 1, 2, .., N,$$

yield the equations of motion (2.15). Thus the system (2.15) defines a Lagrangian correspondence, (or multi-valued mapping), in the sense of Veselov, [**17**]. Choosing canonical momenta via the relations

$$(3.3) \qquad \widetilde{p}_i = \frac{\partial \mathcal{L}}{\partial \widetilde{x}_i} \ , \ p_i = -\frac{\partial \mathcal{L}}{\partial x_i},$$

we can introduce the standard symplectic form $\Omega = \sum_{j=1}^{N} dp_j \wedge dx_j$ on the phase space $\mathcal{M}$ with canonical coordinates $\{x_i, p_i\}_{i=1,...,N}$. Clearly, $\Omega$ is preserved under the correspondence (3.3), $\widetilde{\Omega} = \Omega$.

The above discussions allow us to make the following choice of canonical variables $(P_i, x_i)$ of the $i$th particle

$$P_i = p_i + \sum_{\substack{j=1 \\ j \neq i}}^{N} \frac{1}{x_i - x_j} , \qquad (3.4)$$

leading to the standard Poisson brackets

$$\{P_i, x_j\} = \delta_{ij}, \quad \{P_i, P_j\} = \{x_i, x_j\} = 0 . \qquad (3.5)$$

It follows that in terms of these canonical variables, the Lax matrix $L$ can be written as:

$$L = \sum_{i=1}^{N} P_i e_{ii} - \sum_{\substack{i,j=1 \\ i \neq j}}^{N} \frac{e_{ij}}{x_i - x_j}. \qquad (3.6)$$

Eq. (3.6) is the usual $L$-matrix for the (continnum) rational CM model, so the fundamental Poisson bracket structure between $L$ and $L$ is the same as in the continuum case, i.e.,

$$\{L \overset{\otimes}{,} L\} = [r_{12}, L \otimes 1] - [r_{21}, 1 \otimes L], \qquad (3.7)$$

where the $r$-matrix is given in [**18**].

Thus, in going from the continuous to the discrete CM model, both the $L$-operator as well as the $r$-matrix remain the same. What changes is the $M$-matrix, which will now depend on the canonical variables only implicitly.

The involutivity of the invariants

$$\{I_k, I_l\} = \{Tr(L^k), Tr(L^l)\} = 0 \qquad \text{for all} \quad k, l = 1, 2, .... , \qquad (3.8)$$

that follows as a consequence of the $r$-matrix structure (3.7), will lead to the integrability of the discrete-time model by an argument presented elegantly in [**19**], forming the discrete counterpart of the Arnol'd-Liouville theorem, [**17**]. Therefore, we have proved the complete integrability of the discrete-time CM model.

**3.2. Exact integration.** The integration of the rational discrete CM model (2.15) is performed roughly along the same line as for the continuum model. In the discrete case the initial value problem is posed by asking to find $\{x_i(n)\}$ for given initial data $\{x_i(0)\}$ and $\{x_i(1) \equiv \widetilde{x}_i(0)\}$, where $x_i(n)$ denotes the position of the particles $x_i$ at the $n$th time-step. This is done as follows. Introducing the $N \times N$ matrices

$$X = \sum_{i=1}^{N} x_i e_{ii} , \qquad E = \sum_{i,j=1}^{N} e_{ij}, \qquad (3.9)$$

and using the expressions for $L$ and $M$ in (2.10-2.11), it is readily shown that

$$(\widetilde{L} - M)E = 0 , \qquad (3.10)$$
$$E(L - M) = 0 , \qquad (3.11)$$
$$\widetilde{X}M - MX = -E , \qquad (3.12)$$
$$XL - LX = 1 - E, \qquad (3.13)$$

in addition to (2.13). Now from the Lax equation (2.13) it is clear that we can put $M = \widetilde{U}U^{-1}$ and $L = U\Lambda U^{-1}$, where $U$ is an invertible $N \times N$ matrix, and where the matrix $\Lambda$ is constant, $\widetilde{\Lambda} = \Lambda$, as a consequence of (2.13). Then, from (2.13) and (3.10-3.13) we obtain

$$\widetilde{Y} = \Lambda^{-1}(Y - \Lambda^{-1})\Lambda. \tag{3.14}$$

where $Y \equiv U^{-1}XU$. Noting the invariance of $\Lambda$ under the discrete-time shift we can easily solve eq. (3.14). Further using the relation $\Lambda = U(0)^{-1}L(0)U(0)$, which we calculate from the initial data, we then obtain the following result: *the position coordinates $\{x_i(n)\}$ of the particles at the discrete time $n$, evolving under the discrete equation of motion (2.15), are given by the eigenvalues of the matrix*

$$Y(n) = U(0)^{-1}L(0)^{-n}\left[X(0) - nL(0)^{-1}\right]L(0)^{n}U(0). \tag{3.15}$$

Comparing eq. (3.15) with a similar formula for the continuous case, [**3**], we have to think of $L(0)^{-1}$ as playing the role of the resolvant, noting that we can always extract a constant parameter from $L(0)$ that will become small in the continuum limit. Thus, eq. (3.15) shows that the discrete step $Y(n) \mapsto Y(n+1)$ is actually a time-step 1 update of the continuous flow[1]. ¿From (3.15) we see that at each discrete-time value the positions of the particles is uniquely determined up to a permutation of the particles. One can fix an order of the particles (thus fixing a branch of the correspondence) by imposing–as is usually done in the continuum case– that $x_1(n) < x_2(n) < \cdots < x_N(n)$.

**3.3. Factorization problem.** We can think about the discrete-time CM model from the view point of a factorization problem. First, we note that the matrix $M$ given in (2.11) is actually a Cauchy matrix. Thus, as a consequence of Cauchy's identity, [**21**], its determinants and its minors can be expressed as products of the form

$$\det(M) = \frac{\prod_{k<\ell}(x_k - x_\ell)(\widetilde{x}_\ell - \widetilde{x}_k)}{\prod_{k,\ell}(x_k - \widetilde{x}_\ell)}. \tag{3.16}$$

Furthermore, we see that from (2.13) and (3.10-3.13), we can obtain

$$(LM^{-1})\widetilde{X} - X(LM^{-1}) = -M^{-1}, \tag{3.17}$$

which tell us that

$$(LM^{-1})_{ij} = \frac{(M^{-1})_{ij}}{x_i - \widetilde{x}_j}, \tag{3.18}$$

whose entries can be expressed as products as a consequence of the Cauchy's identity. Then, we can think of the discrete-time flow as coming from a factorization in terms of a Cauchy matrix $M$ and a "Cauchy-like" matrix $K = LM^{-1}$, whose entries factorize into products. Thus we can write down a dressing chain as follows:

$$L = KM \mapsto \widetilde{L} = MK = \widetilde{K}\widetilde{M} \mapsto \ldots.$$

---

[1] We note here that the relevant expression given in our previous paper, i.e., the eq. (33) in [**11**], is equivalent to the above expression (3.15), which can be proved by using (3.13). A similar expression to eq. (3.15) can be obtained for the mastersymmetries of the discrete model, cf. [**20**].

Furthermore, the matrix $M$ can be indeed regarded as a dressing matrix, as it can be shown that multiple iterates of the form

$$M(n+m)M(n+m-1)\ldots M(n+1)M(n)$$

depend only on the initial and final time-variables $n+m$ and $n$, but not on the values of the time-variable in the intermediate steps.

**3.4. Continuum limit.** In a continuum limit (2.15) goes to that of the original (rational) CM model, which can be seen as follows. We first set

$$(3.19) \qquad x_i = z_i + n\Delta, \quad i = 1, 2, ..., N,$$

where $\Delta$ is a small (constant) parameter and

$$(3.20) \qquad \widetilde{x}_i = \widetilde{z}_i + (n+1)\Delta, \quad i = 1, 2, ..., N.$$

Then, in the continuum limit, we write

$$(3.21) \qquad \widetilde{z}_i = z_i + \epsilon \dot{z}_i + 1/2\epsilon^2 \ddot{z}_i + ...,$$

for all $i = 1, 2, .., N$, where $\dot{z}_i \equiv \frac{d}{dt}z_i$ and $\epsilon$ is the time-step parameter which, we suppose, is in the order of $O(\Delta^2)$. Substituting (3.19)- (3.21) into (2.15), we get as the leading order term of (2.15), i.e.

$$(3.22) \qquad \ddot{z}_i = -2g \sum_{\substack{j=1 \\ j \neq i}}^{N} \frac{1}{(z_i - z_j)^3}, \quad i = 1, 2, ..., N,$$

where $g \equiv \Delta^4/\epsilon^2$. It is clear that (3.22) are exactly the equations of motion of the (continuous) rational CM model. It is interesting to note that the coupling constant of the continuous model arises from the discrete-time step.

## 4. Elliptic generalization

**4.1. Elliptic discrete-time CM model.** Now we generlize the above result to the elliptic case, and we are to construct an (integrable) elliptic discrete-time CM model, which goes exactly to (1.1) in a continuum limit. In order to do that, let us first recall the basic definitions and some properties of the classical functions of Weierstrass, see e.g. [22]. Let $\omega_1$ and $\omega_2$ be a pair of periods. The Weierstrass sigma-function is an entire function defined by

$$(4.1) \qquad \sigma(x) = x \prod_{k,l \neq 0} (1 - \frac{x}{\omega_{kl}}) \exp\left[\frac{x}{\omega_{kl}} + \frac{1}{2}(\frac{x}{\omega_{kl}})^2\right],$$

with $\omega_{kl} = k\omega_1 + l\omega_2$. The remaining functions can be defined by the relations

$$(4.2) \qquad \zeta(x) = \frac{\sigma'(x)}{\sigma(x)}, \quad \wp(x) = -\zeta'(x).$$

Note that both the zeta-function and the sigma-function are odd functions and they satisfy the following fundamental relation

$$(4.3) \qquad \zeta(\lambda) + \zeta(\mu) + \zeta(\nu) - \zeta(\lambda + \mu + \nu) = \frac{\sigma(\lambda + \mu)\sigma(\mu + \nu)\sigma(\nu + \lambda)}{\sigma(\lambda)\sigma(\mu)\sigma(\nu)\sigma(\lambda + \mu + \nu)}.$$

8     F. W. NIJHOFF AND G. D. PANGNow by using (4.3), it can be directly checked thatNow by using (4.3), it can be directly checked that

$$(4.4) \qquad u = \zeta(x) \quad \text{with} \quad x = \xi + n\alpha + m\beta$$

is a simple elliptic solution to the semi-discretized version of KP equation (2.1), where $p = \zeta(\alpha)$ and $q = \zeta(\beta)$, corresponding to the following solution of the linear equation (2.2)

$$(4.5) \qquad \varphi(x;\kappa) = \frac{\sigma(x-\kappa)}{\sigma(x)\sigma(\kappa)} \left(\frac{\sigma(\alpha+\kappa)}{\sigma(\alpha)\sigma(\kappa)}\right)^n \left(\frac{\sigma(\beta+\kappa)}{\sigma(\beta)\sigma(\kappa)}\right)^m e^{\zeta(\kappa)\xi}.$$

Similar to the rational case, we suppose that

$$(4.6) \qquad u = \sum_{i=1}^N \zeta(\xi - x_i),$$

$$(4.7) \qquad \varphi = \sum_{i=1}^N b_i \Phi_\kappa(\xi - x_i) e^{\zeta(\kappa)\xi},$$

where

$$(4.8) \qquad \Phi_\kappa(x) \equiv \frac{\sigma(x-\kappa)}{\sigma(x)\sigma(\kappa)}.$$

Substituting (4.6-4.7) into (2.2) and equating to zero the coefficients of $\Phi_\kappa(\xi - x_i)$ and $\Phi_\kappa(\xi - \widetilde{x}_i)$, we obtain

$$(4.9) \qquad pb = Lb,$$
$$(4.10) \qquad \widetilde{b} = Mb,$$

where the vector $b = (b_1, b_2, ..., b_N)^T$ and the matrices

$$(4.11) \quad L = \sum_{i=1}^N \left( \sum_{j=1}^N \zeta(x_i - \widetilde{x}_j) - \sum_{\substack{j=1 \\ j \neq i}}^N \zeta(x_i - x_j) \right) e_{ii} - \sum_{\substack{i,j=1 \\ i \neq j}}^N \Phi_\kappa(x_i - x_j) e_{ij},$$

$$(4.12) \quad M = \sum_{i,j=1}^N \Phi_\kappa(\widetilde{x}_i - x_j) e_{ij}.$$

The compatibility of (4.9-4.10), i.e., $\widetilde{L}M = ML$ give the equations of motion of the elliptic discrete-time CM model:

$$(4.13)$$
$$\zeta(x_i - \widetilde{x}_i) + \zeta(x_i - \underset{\sim}{x}_i) + \sum_{\substack{j=1 \\ j \neq i}}^N \left( \zeta(x_i - \widetilde{x}_j) + \zeta(x_i - \underset{\sim}{x}_j) - 2\zeta(x_i - x_j) \right) = 0,$$
$$i = 1, 2, ..., N.$$

Eq. (4.13) defines an elliptic correspondence, which below is shown to be integrable along the same lines as the rational CM model. Apart from mappings that can be inferred from a lattice version of the Landau-Lifshitz equations that was derived some years ago in [23], this is one of the first examples of an integrable elliptic mapping. We remark that the elliptic case was curiously absent in the list of integrable mappings derived by Suris in [24]. Note that in the degenerate case when $\omega_1 \to \infty$ and $\omega_2 \to i\infty$, we recover from eq. (4.13) the rational discrete-time CM model (2.15). Obviously, in the intermediate degenerate limits with only one of the periods $\omega_{1,2}$ taking an infinite value, it goes to the trigonometric and hyperbolic cases.

**4.2. Complete integrability.** The complete integrabilty of the elliptic discrete-time CM model (4.13) can be proved in a similar way to that of the rational case. We only note here that in the elliptic case the action is

$$(4.14) \quad S = \sum_n \mathcal{L}(x, \widetilde{x}) = \sum_n \left( -\sum_{i,j=1}^N \log|\sigma(x_i - \widetilde{x}_j)| + \sum_{\substack{i,j=1\\i\neq j}}^N \log|\sigma(x_i - x_j)| \right),$$

and the canonical momenta $P_i$ in the standard Poisson brackets (3.5) now is given by

$$(4.15) \quad P_i = -\frac{\partial \mathcal{L}}{\partial x_i} + \sum_{\substack{j=1\\j\neq i}}^N \zeta(x_i - x_j) = \sum_{j=1}^N \zeta(x_i - \widetilde{x}_j) - \sum_{\substack{j=1\\j\neq i}}^N \zeta(x_i - x_j).$$

In terms of the canonical variables, the Lax matrix $L$ in (4.11) has again the same form as that of the continuum (elliptic) CM model. Thus we can use the $r$-matrix structure given in [25] to establish the involutivity of the invariants, i.e.,

$$(4.16) \quad \{I_k, I_l\} = \{Tr(L^k), Tr(L^l)\} = 0 \quad \text{for all} \quad k, l = 1, 2, \ldots,$$

where $L$ is given by (4.11).

**4.3. Continuum limit.** It can be shown that in the continuum limit described in section 3.4 the leading order term of (4.13) is exactly (1.1), where the $g$ is equal to $-\Delta^2/[\epsilon^2 \zeta'(\Delta)]$.

## 5. Lattice KP equation

In the previous sections we use pole-expansion for a semi-continuous version of the KP equation. In this section we will show that similar pole-expansions can be also done for the fully discretized KP equation, i.e., the lattice KP equation presented in [16], and the discrete-time CM model can be also obtained from pole-solutions of the lattice KP equation.

The lattice KP equation reads

$$(5.1) \quad \frac{p - r + \widehat{u}' - \widetilde{\widehat{u}}}{p - r + u' - \widetilde{u}} = \frac{q - r + \widetilde{u}' - \widetilde{\widehat{u}}}{q - r + u' - \widehat{u}},$$

where $r$ is the third (lattice) parameter in addition to $p$ and $q$. The $'$ in (5.1) denotes a third shift or translation. Note that the three lattice directions are on the same footing. Eq. (5.1) is the compatibility condition of the following three equations, forming – as it were – an inhomogeneous linear system for (5.1),

$$(5.2) \quad \widetilde{\varphi} = (p - \widetilde{u})\varphi + \psi,$$
$$(5.3) \quad \widehat{\varphi} = (q - \widehat{u})\varphi + \psi,$$
$$(5.4) \quad \varphi' = (r - u')\varphi + \psi,$$

in which $\psi$ is any function independent of the particular discrete direction. Thus eq. (5.2-5.4) can be considered to be an *inhomogeneous* Lax representation of (5.1).

Again, direct calculation shows that

$$(5.5) \qquad u = \zeta(x) \quad \text{with} \quad x = \xi + n\alpha + m\beta + h\gamma$$

is a simple elliptic solution to the lattice KP equation (5.1), where $p = \zeta(\alpha)$, $q = \zeta(\beta)$ and $r = \zeta(\gamma)$, corresponding to the following solution of the Lax 'triple'

$$(5.6) \quad \varphi(x;\kappa) = \frac{\sigma(x-\kappa)}{\sigma(x)\sigma(\kappa)} \left(\frac{\sigma(\alpha+\kappa)}{\sigma(\alpha)\sigma(\kappa)}\right)^n \left(\frac{\sigma(\beta+\kappa)}{\sigma(\beta)\sigma(\kappa)}\right)^m \left(\frac{\sigma(\gamma+\kappa)}{\sigma(\gamma)\sigma(\kappa)}\right)^h ,$$
$$\psi(x;\kappa) = [\zeta(\kappa) + \zeta(x-\kappa)] \varphi(x;\kappa) ,$$

where $h$ is an integer similar to $n$ and $m$ and increased by value 1 when operated by $'$. Here and hereafter we always assume $\xi$ is a dummy variable.

Similarly if we suppose that $u$ and $\varphi$ are given by (4.6-4.7), and

$$(5.7) \qquad \psi = \sum_{i=1}^{N} b_i[\zeta(\kappa) + \zeta(\xi - x_i - \kappa)]\Phi_\kappa(\xi - x_i)e^{\zeta(\kappa)\xi}$$
$$+ \sum_{\substack{i,j=1 \\ i \neq j}}^{N} b_j \zeta(\xi - x_i)\Phi_\kappa(\xi - x_j)e^{\zeta(\kappa)\xi},$$

then from (5.2) we can obtain again the elliptic discrete-time CM model (4.13) and its Lax representation (4.9-4.10).

Of course, from the other two linear equations (5.3-5.4), we can get equations similar to (4.13), which represent the other two flows in the other two lattice directions. These three flows are, in fact, all compatible if

$$(5.8) \qquad p - q = \sum_{j=1}^{N} \zeta(x_i - \widetilde{x}_j) - \sum_{j=1}^{N} \zeta(x_i - \hat{x}_j),$$

$$(5.9) \qquad q - r = \sum_{j=1}^{N} \zeta(x_i - \hat{x}_j) - \sum_{j=1}^{N} \zeta(x_i - x_j{'}),$$

for all $i = 1, 2, ..., N$. More details about these compatible conditions and elliptic solutions of the lattice KP equations will be published in [**26**].

## 6. Discussion

In this paper, we have constructed a discrete-time CM model. This discrete model is also integrable and is the iterate of a symplectic correspondence. The original CM model is a limit of this discrete-time CM model, and in this limit the coupling constant for the long-range interaction term is encoded in the discrete-time step parameter. The result of the present paper fills a gap in a series of various deformations and generalizations of the CM model (such as the quantization, the deformation to the relativistic case or the generalization to different root systems), and it invites a number of interesting problems to be studied. First of all, one may ask whether one can find time-discretizations of the relativistic version of the CM model, [**4**], cf. also [**27**]. In fact, Suris, in [**28**], has found an interesting connection between the discrete-time Toda model and its relativistic version. Furthermore, a direct connection exists between the sine-Gordon soliton solutions and the relativistic CM model, cf. [**29, 4**]. This could lead to an alternative way to discretize the model. Secondly, for the discrete-time model, following the similarities with the structure of the integrable quantum mappings studied in [**30**], one should investigate not only the $L$-part of the Lax pair, but also take the $M$-part under consideration in the $r$-matrix structure. This has been done for the integration of mappings of KdV type in [**31**]. Finally, these investigations should also be pursued on the quantum level. Although, the discrete-time model has an obvious quantum counterpart, in much the same way as the continuum model, the work of [**30**] on quantum mappings, as well as the work [**32**] on the quantum CM model, indicate that some important modification (e.g. with respect to the construction of exact quantum invariants of the discrete-time flow) might be expected. The quantum model in itself might shed some new light on the random matrix approach to quantum chaos, as –in principle– it provides us with a finite and exact algorithm to calculate multi-particle propagators in the long-range model, in the spirit of [**33**]. All these problems are being investigated and will be discussed in detail in future publications.

DEPARTMENT OF MATHEMATICS, UNIVERSITY OF PADERBORN, D-33095 PADERBORN, GERMANY

   *E-mail address*: nijhoff@uni-paderborn.de

INSTITUTE OF PHYSICS, CHINESE ACADEMY OF SCIENCES, BEIJING, CHINA

   *Current address*: Department of Mathematics, University of Paderborn, D-33095 Paderborn, Germany

   *E-mail address*: pang@uni-paderborn.de